\documentclass[11pt,twoside]{article}
\usepackage{asp2010}
\usepackage{german, ngerman,natbib}
\usepackage[german]{babel}

\resetcounters

\begin{document}
\title{New HW Virginis Systems from the MUCHFUSS project}
\author{Veronika Schaffenroth$^1$$^{,2}$, Stephan Geier$^3$$^{,1}$, Ingrid Barbu-Barna$^1$, Uli Heber$^1$, Thomas Kupfer$^4$ and Oliver Cordes$^5$
\affil{$^1$Dr Karl Remeis Observatory and ECAP, Astronomical Institute, Friedrich-Alexander University Erlangen-Nuremberg, Sternwartstrasse 7, 96049 Bamberg, Germany}
\affil{$^2$Institute for Astrophysics and Particle Physics, University of Innsbruck, Technikerstrasse 25/8, 6020 Innsbruck, Austria}
\affil{$^3$European Southern Observatory, Karl-Schwarzschild-Str. 2, 85748 Garching, Germany}
\affil{$^4$Department of Astrophysics/IMAPP, Radboud University Nijmegen, P.O. Box
9010, 6500 GL Nijmegen, The Netherlands}
\affil{$^5$Argelander-Institut f\"ur Astronomie, Auf dem H\"ugel 71, D-53121 Bonn, Germany}}

\begin{abstract}
In course of the MUCHFUSS project we found two new HW Virginis systems. J192059+372220 (J1920) has an orbital period of 0.169d and an M dwarf companion with a mass of 0.116 $M_{\odot}$. J162256+473051 (J1622) has the shortest period (0.069 d) of all known HW Virginis systems. Its companion has a mass of 0.0635$\pm$ 0.004 $M_{\odot}$ and is therefore the second close brown dwarf found orbiting a subdwarf B star.
\end{abstract}

\section{Introduction}
HW Virginis systems are eclipsing subdwarf binaries with cool, low-mass companion. They are a very important group among the subdwarf binaries, as a combined spectroscopic and lightcurve analysis can determine the masses and radii of both stars. They are easily recognized from their very characteristic lightcurve with the prominent reflection effect resulting from the large temperature difference of both stars. The hemisphere of the cool companion facing the hot subdwarf is heated up. Therefore the contribution of the companion to the light varies with the orbital phase and reaches its maximum shortly before the secondary eclipse, where almost the complete hot side of the companion can be seen, and its minimum in the primary eclipse, where only the cool side is visible.\\ Based on the lightcurve it is possible to distinguish between a low mass main sequence companion and a white dwarf, as a white dwarf does not show a reflection effect because of its small radius. Therefore we started a photometric follow-up campaign of the hot subdwarf binaries selected in the course of the MUCHFUSS project (\citet{2011A&A...530A..28G}; Geier et al. these proceedings). For the photometric follow-up we used the Bonn Simultaneous Camera (BUSCA, see \citet{1999SPIE.3649..109R}) at the 2.2m telescope at the Calar Alto, which can observe in 4 bands simultaneously. We did not use any filters but instead the intrinsic transmission curve given by the beam splitter.
\section{Data analysis}
For the spectroscopic analysis we used medium resolution spectra. To determine the radial velocity, a fit of Gaussians, Lorentzians and polynomials to the Balmer and helium lines was performed. To derive the radial velocity curve, sine curves were fitted to the radial velocities. The atmospheric parameters are determined by a fit of a grid of synthetic spectra, that was calculated using LTE model atmospheres with solar metallicity and metal line blanketing \citep{heber00}, to the Balmer and helium lines.\\
For the lightcurve analysis we used MORO (see \citet{drechsel:1995} and \citet{vs} for more information). As the lightcurve analysis involves 12 + 5n (n is the number of lightcurves) not independent parameters, a degeneracy in the solutions can be seen and a single solution can not be determined. Therefore it is necessary to constrain as many parameters as possible from the spectroscopic analysis. Moreover, several solutions are calculated for different fixed mass ratios, as the mass ratio can not be determined because of the degeneracy.
\section{The sdB+M binary J1920}
\begin{figure}
\begin{minipage}{0.45\linewidth}
\includegraphics[width=1.0\linewidth]{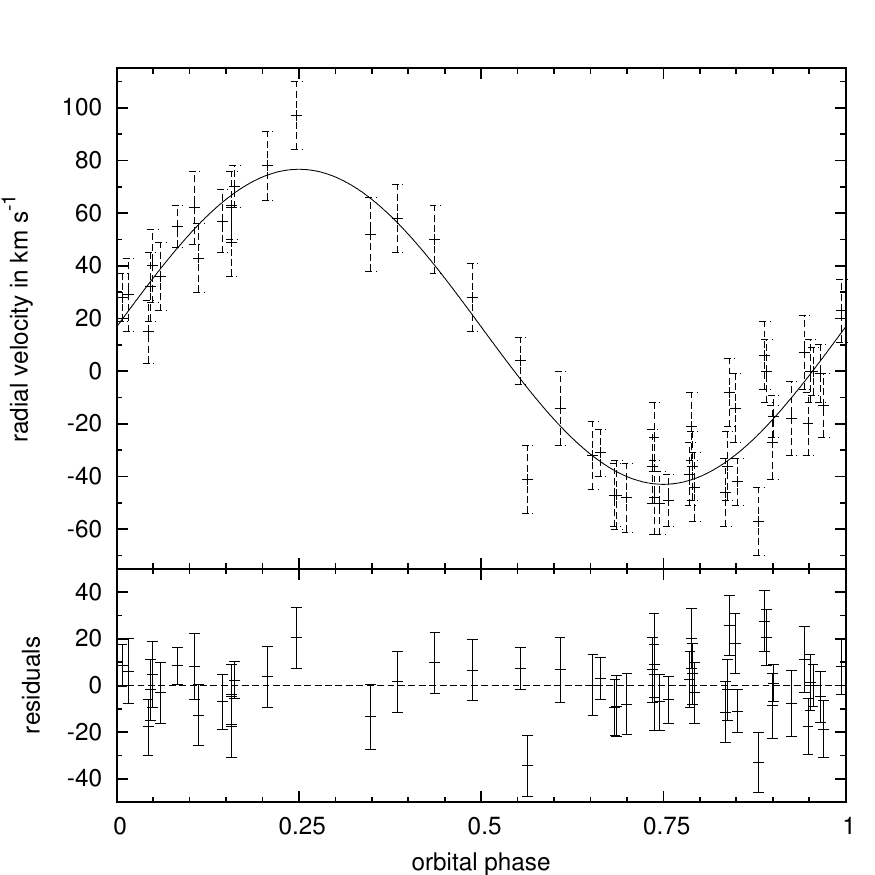}
\end{minipage}
\begin{minipage}{0.55\linewidth}
\includegraphics[angle=-90,width=1.0\linewidth]{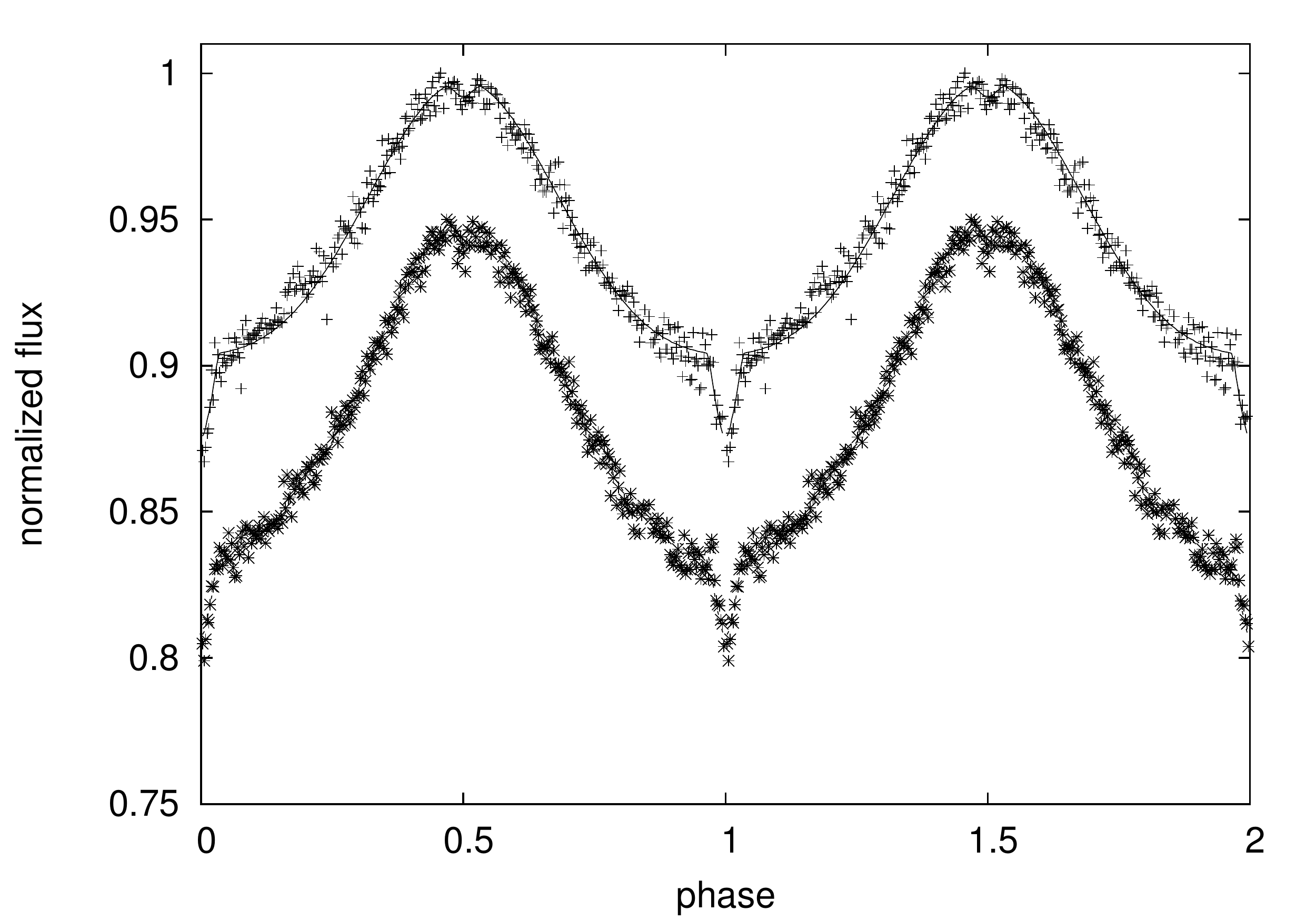}
\end{minipage}
\caption{In the left panel the RV curve of J1920 is displayed. The solid line represents the best fit of a sine curve to the data. In the right panel the lightcurve of J1920 in the $B_B$, and $V_B$ band is illustrated together with the lightcurve solution for a canonical sdB mass.}
\label{1920}
\end{figure}

J1920 was selected as a high priority target for the MUCHFUSS project as it shows a significant radial velocity shift in the four SDSS spectra, which were obtained within 1.5h. Therefore we observed a 1.75 h lightcurve with BUSCA, which showed a reflection effect. Another lightcurve of the complete orbit revealed the grazing eclipses. Hence we declared it a high priority target for spectroscopic follow-up and obtained 39 spectra with TWIN at the 3.5m telescope at the Calar Alto and 11 spectra with ISIS at the William Herschel Telescope at the Roques de los Muchachos Observatory in La Palma.\\
The radial velocity curve, which covers the whole orbit nicely, is shown in Fig.\ref{1920}. It gives a semiamplitude of 59.8 $\pm$ 2.5 $\rm km\,s^{-1}$ and a period of 0.169 d. The lightcurves in $B_B$ and $B_V$ with the best fit are also shown in Fig. \ref{1920}. We used only two of the four lightcurves as the S/N in the other two bands is not high enough. Unfortunately it is not possible to determine the period of the system from the lightcurve as we have only one full orbit. Until now only a solution for the canonical sdB mass (0.47 $M_{\odot}$) is being calculated. The results of the spectroscopic and photometric analysis are compiled in Table \ref{result_1922} The companion is a late M dwarf with a mass of 0.116 $M_{\odot}$, if the sdB has the canonical mass.
	\begin{table}\centering
	\begin{tabular}{|c| c| c c c|}
		\hline\hline
		K & $\rm [km\,s^{-1}]$ & 59.8 & $\pm$ & 2.5\\
		$\gamma$ & $\rm [km\,s^{-1}]$ &16.8  & $\pm$ & 2.0\\
		P & d & 0.168876 & $\pm$ & 0.00035\\
		a & $\rm R_{\odot}$ & 1.078 & $\pm$ & 0.0449\\\hline
		$\rm T_{eff}$& K& 27500 &$\pm$ &1000\\
		$\log{g}$& &5.4&$\pm$&0.1\\\hline
		i & $^{\circ}$  & 67.2 &  & \\
		$\rm M_{sdB}$ & $\rm M_{\odot}$ &0.47 &  & \\
		$\rm M_{comp}$ & $\rm M_{\odot}$ & 0.116 & $\pm$ & 0.007\\
		$\rm r_{sdB}$ & $\rm R_{\odot}$ & 0.287 & $\pm$ & 0.010\\
		$\rm r_{comp}$ & $\rm R_{\odot}$ & 0.190 & $\pm$ & 0.008\\
		$\rm \log{g} $ &   & 5.35 & $\pm$ &0.014 \\
		\hline\hline
	\end{tabular}
	\caption{Results of the spectroscopic and lightcurve analysis for J1920}
	\label{result_1922}
	\end{table}

\section{The sdB+BD binary J1622}
\begin{figure}
\begin{minipage}{0.45\linewidth}
\includegraphics[width=1.0\linewidth]{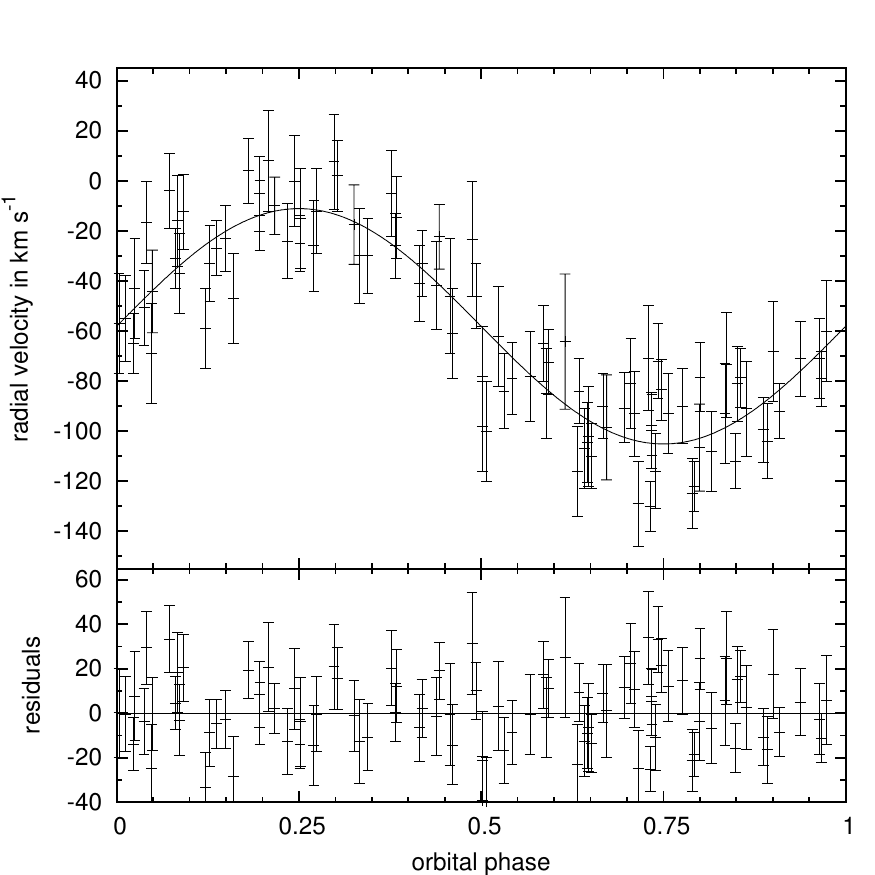}
\end{minipage}
\begin{minipage}{0.55\linewidth}
\includegraphics[angle=-90,width=1.0\linewidth]{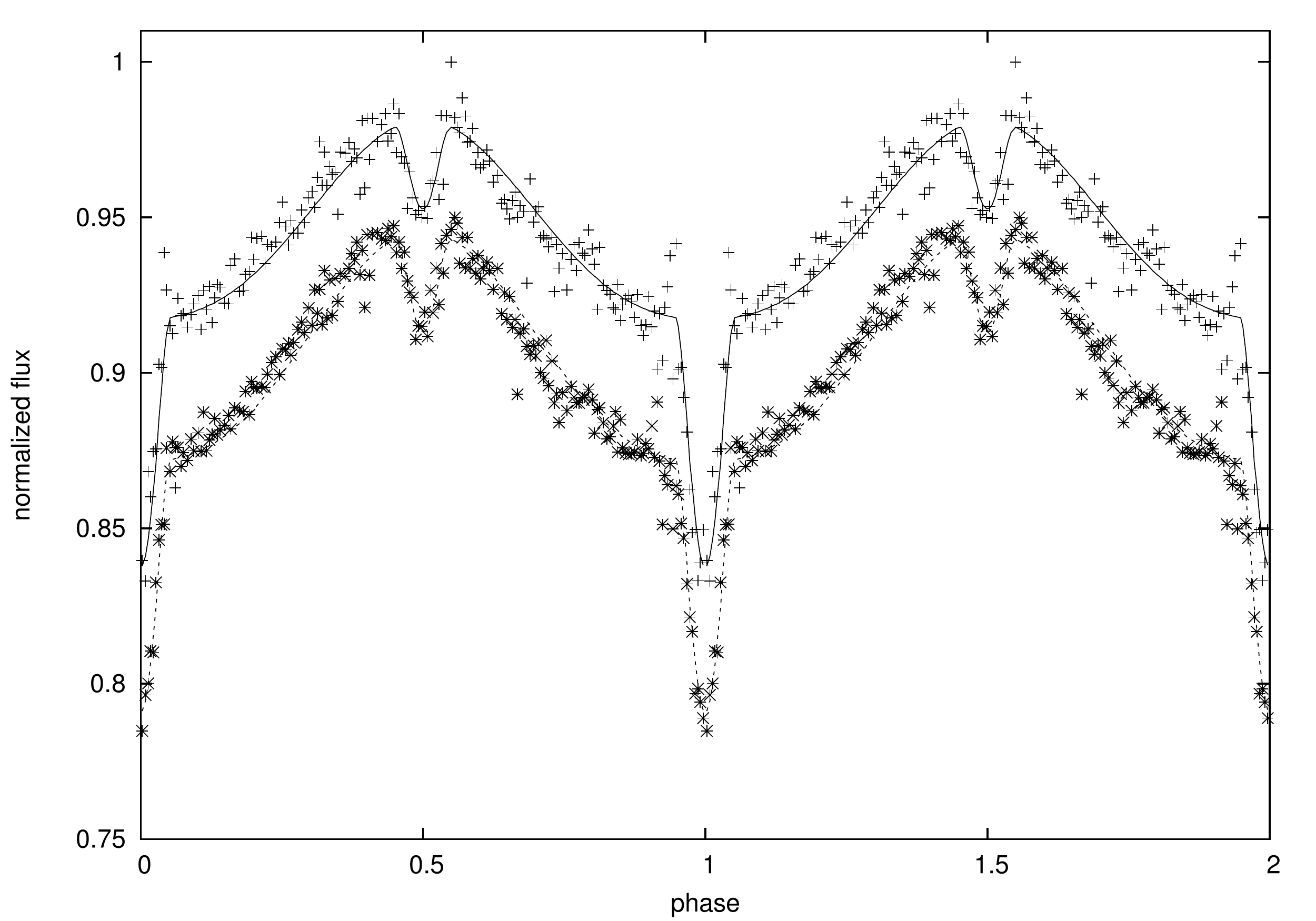}
\end{minipage}
\caption{In the left panel the RV curve of J1622 is displayed. The solid line represents the best fit of a sine curve to the data. In the right panel the lightcurve of J1622 in the $B_B$, and $V_B$ band is illustrated together with the most probable lightcurve solution with a canonical sdB mass.}
\label{1622}
\end{figure}

\begin{figure}
\includegraphics[width=0.9\linewidth]{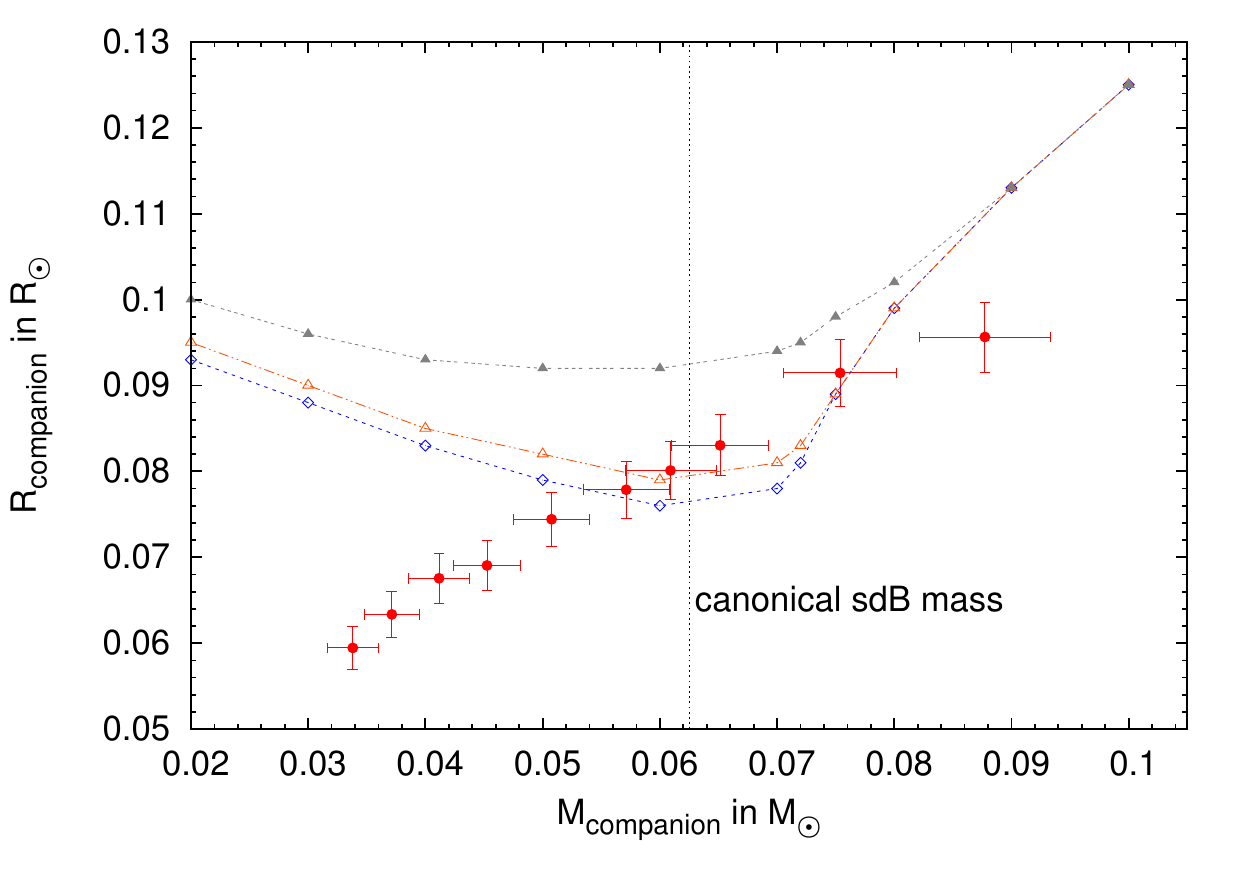}\\
\caption{Comparison of theoretical mass-radius-relation of a brown dwarf by \citet{baraffe:2003} for an age of 1 Gy (filled triangles), 5 Gy (triangles) and 10 Gy (diamond) to result from the lightcure analysis}
\label{1622_bd}
\end{figure}

J1622 got a very high priority, as the SDSS spectra showed a high shift in a short time. Therefore it was selected for a spectroscopic follow-up. The RV-curve had a very short period of only 0.069 d. This made it a very high priority target for a photometric follow-up. The lightcurves showed the characteristic reflection effect and eclipses. Therefore more medium resolution spectra were obtained with TWIN and also higher resolution spectra with ESI/Keck to determine the rotational velocity. In Fig. \ref{1622} you can see the radial velocity curve and the lightcurve with the fit of the best solution. As the degeneracy does not allow us to determine a single solution, another criterion is needed to find the correct solution. Therefore we use the theoretical mass-radius relation for the companion, which can be found in Fig. \ref{1622_bd}. The comparison with the mass and radius from the lightcurve analysis for different mass ratios shows an agreement only for masses of the sdB of $\sim$0.47 and 0.63 $M_{\odot}$. Hence the most probable solution is an sdB mass of 0.47 $M_{\odot}$ and a mass of the companion of 0.0635 $\pm$ 0.004 $M_{\odot}$, which is lower than the limiting mass for core hydrogen-burning. The companion is therefore a brown dwarf, only the second one found in close orbit around an sdB star (see \citet{2011A&A...530A..28G}). For  hot Jupiters in short orbits an inflation of the planet is observed. If we assume an inflation of the brown dwarf no agreement between the theoretical and the measured mass-radius relation can be found and we can consequently exclude an inflation of the brown dwarf. In table \ref{result_1622} you can find all the results of the spectroscopic and photometric analysis. From the ESI spectra (R=8000) the projected rotational velocity of the sdB primary has been measured from the line broadening ($\rm v_{rot} \sin{i}$ = 66 km/s). It is only about half the rotational velocity expected for a synchronized rotation of the sdB with its orbital motion. Despite its very short period, we therefore found the sdB primary to rotate non-synchronously (see also Kawaler et al. these proceedings). 
	\begin{table}\centering
	\begin{tabular}{|c| c |c c c|}
		\hline\hline
		K & $\rm [km\,s^{-1}]$ & 47.0 & $\pm$ & 2.0\\
		$\gamma$ & $\rm [km\,s^{-1}]$ & -58.1 & $\pm$ & 1.5\\
		P & d & 0.069792 & $\pm$ & 0.000045\\
		a & $\rm R_{\odot}$ & 0.5707 & $\pm$ & 0.0243\\\hline
		$\rm T_{eff}$& K& 29000 &$\pm$ &1000\\
		$\log{g}$& &5.65&$\pm$&0.1\\\hline
		i & $^{\circ}$  & 72.22 & $\pm$&1.16  \\
		$\rm M_{sdB}$ & $\rm M_{\odot}$ &0.479 & $\pm$ & 0.03\\
		$\rm M_{comp}$ & $\rm M_{\odot}$ & 0.0635 & $\pm$ & 0.004\\
		$\rm r_{sdB}$ & $\rm R_{\odot}$ & 0.168 & $\pm$ & 0.007\\
		$\rm r_{comp}$ & $\rm R_{\odot}$ & 0.085 & $\pm$ & 0.003\\
		$\rm \log{g} $&   & 5.67 & $\pm$ &0.019 \\\hline
		$\rm v_{rot}$&$\rm [km\,s^{-1}]$&66&$\pm$&7\\
		\hline\hline
	\end{tabular}
	\caption{Results of the spectroscopic and lightcurve analysis for J1622}
	\label{result_1622}
	\end{table}
\section{Conclusions}
\begin{figure} 
\includegraphics[width=0.8\textwidth]{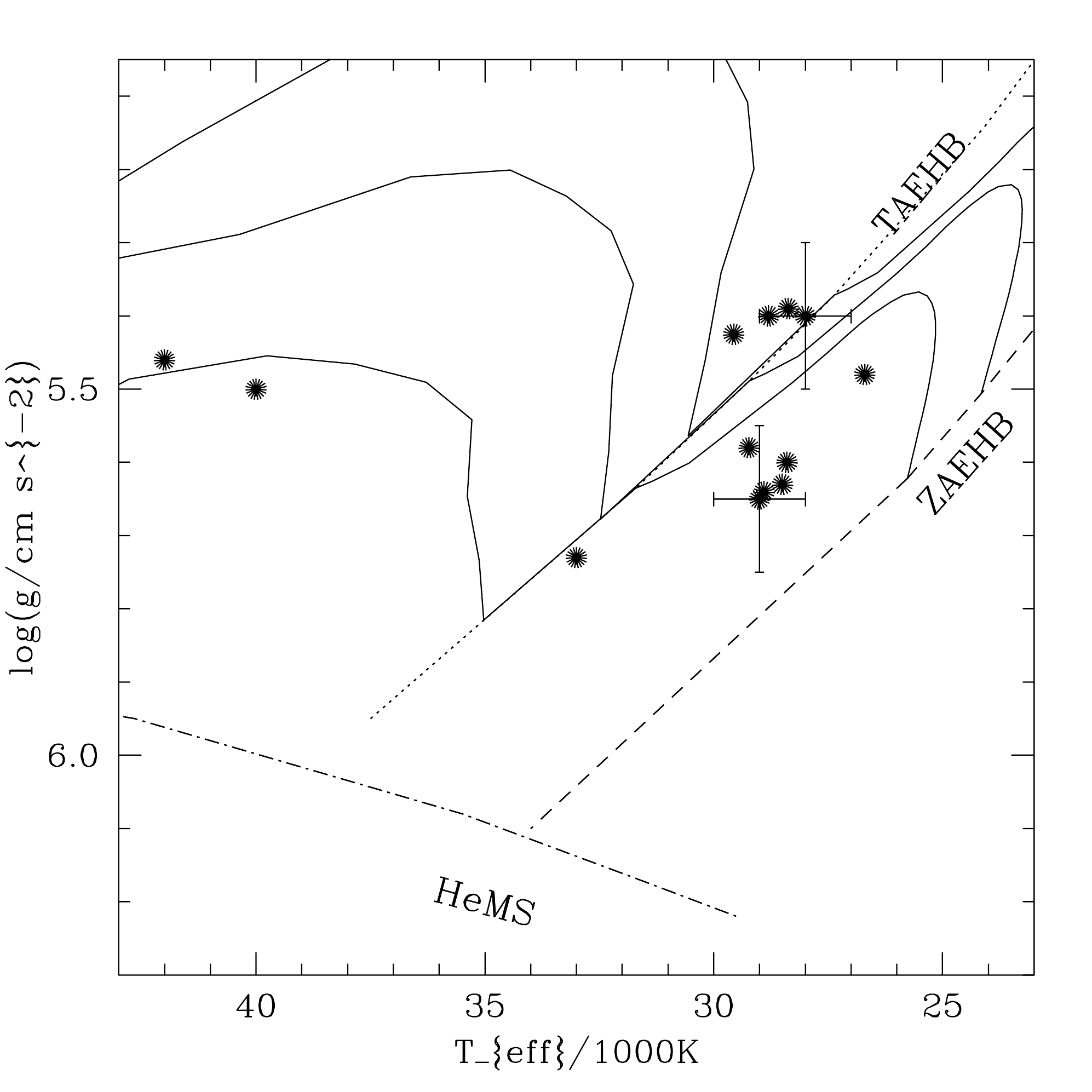}
\caption{$\rm T_{eff}$-log g diagram of the HW Virgins systems. The solid lines are evolutionary tracks by \citet{dorman} for an sdB mass of 0.471, 0.473, and 0.475 $\rm M_{\odot}$ }
\label{teff-logg}
\end{figure}

Fig. \ref{teff-logg} shows the position of the HW Virginis systems in the $\rm T_{eff}-\log{g}$ diagram. Our two new systems have typical parameters for an HW Virginis system. J1920 is more evolved and seems to leave the horizontal branch. It is obvious that all the HW Virginis systems with two exceptions seem to have very similar atmospheric parameters. Normally the HW Virginis systems are found in photometric surveys due to there characteristic lightcurve. We pursue a different approach. As we are looking for HW Virginis systems in a sample of radial velocity variable sdB stars, it is possible for us to determine the fraction of reflection effect binaries in the short period sdB binaries. 66 sdB binaries have been observed, 4 of them show a reflection effect and/or eclipses. This means about 6\% of the companions are low mass main sequence stars. Two of the companions are brown dwarfs, which means that we find about 3\% of the sdB binaries have substellar companions (see also Geier et al. these proceedings).

\begin{thebibliography}{}
\bibitem[Baraffe et al.(2003)]{baraffe:2003} Baraffe, I., Chabrier, G.; Barman, T.~S., Allard, F., Hauschildt, P.~H., A\&A, 402, 701 (2003).
\bibitem[Dorman et al.(1993)]{dorman} Dorman, B., Rood, R.~T., O'Connell, R.~W., APJ,  419, 596 (1993)
\bibitem[Drechsel et al.(1995)]{drechsel:1995} Drechsel, H., Haas, S., Lorenz, R., Gayler, S., A\&A, 294, 723 (1995).
\bibitem[Geier et al.(2011)]{2011A&A...530A..28G} Geier, S., Hirsch, H., Tillich, A., et al.\ 2011, \aap, 530, A28
\bibitem[Geier et al.(2011)]{2011ApJ...731L..22G} Geier, S., Schaffenroth, V., Drechsel, H., et al.\ 2011, \apjl, 731, L22
\bibitem[Heber et al.(2000)]{heber00} Heber, U., Reid, I.~N., \& Werner, K.\ 2000, \aap, 363, 198
\bibitem[Reif et al.(1999)]{1999SPIE.3649..109R} Reif, K., Bagschik, K., de Boer, K.~S., et al.\ 1999, Proc. SPIE, 3649, 109
\bibitem[Schaffenroth et al.(2013)]{vs} Schaffenroth,V. , Geier, S., Heber, U., et al.\ 2013, \aap, 553, A18
\end{thebibliography}

\end{document}